\documentclass[twocolumn,amsmath,aps,fleqn]{revtex4}
\usepackage{graphicx,amssymb}
\begin{document}
\newcommand{\be}{\begin{equation}}
\newcommand{\ee}{\end{equation}}
\newcommand{\bq}{\begin{eqnarray}}
\newcommand{\eq}{\end{eqnarray}}
\newcommand{\bsq}{\begin{subequations}}
\newcommand{\esq}{\end{subequations}}
\newcommand{\bc}{\begin{center}}
\newcommand{\ec}{\end{center}}
\newcommand {\R}{{\mathcal R}}
\newcommand{\al}{\alpha}
\newcommand\lsim{\mathrel{\rlap{\lower4pt\hbox{\hskip1pt$\sim$}}
    \raise1pt\hbox{$<$}}}
\newcommand\gsim{\mathrel{\rlap{\lower4pt\hbox{\hskip1pt$\sim$}}
    \raise1pt\hbox{$>$}}}

\title{Avoiding unrealistic priors: the case of dark energy constraints from the time variation of the fine-structure constant}

\author{P.P. Avelino}
\email[Electronic address: ]{pedro.avelino@astro.up.pt}
\affiliation{Instituto de Astrof\'{\i}sica e Ci\^encias do Espa{\c c}o, Universidade do Porto, CAUP, Rua das Estrelas, PT4150-762 Porto, Portugal}
\affiliation{Centro de Astrof\'{\i}sica da Universidade do Porto, Rua das Estrelas, PT4150-762 Porto, Portugal}
\affiliation{Departamento de F\'{\i}sica e Astronomia, Faculdade de Ci\^encias, Universidade do Porto, Rua do Campo Alegre 687, PT4169-007 Porto, Portugal}

\date{\today}
\begin{abstract}

We critically assess recent claims suggesting that upper limits on the time variation of the fine-structure constant tightly constrain the coupling of a dark energy scalar field to the electromagnetic sector, and, indirectly, the violation of the weak equivalence principle. We show that such constraints depend crucially on the assumed priors, even if the dark energy was described by a dynamical scalar field with a constant equation of state parameter $w$ linearly coupled to the electromagnetic sector through a dimensionless coupling $\zeta$. We find that, although local atomic clock tests, as well as other terrestrial, astrophysical and cosmological data, put stringent bounds on $|\zeta| {\sqrt {|w+1|}}$, the time variation of the fine-structure constant cannot be used to set or to improve upper limits on $|\zeta|$ or $|w+1|$ without specifying priors, consistent but not favoured by current data, which strongly disfavour low values of $|w+1|$ or $|\zeta|$, respectively.  We briefly discuss how this might change with a new generation of high-resolution ultra-stable spectrographs, such as ESPRESSO and ELT-HIRES, in combination with forthcoming missions to map the geometry of the Universe, such as Euclid, or to test the equivalence principle, such as MICROSCOPE or STEP. 

\end{abstract}
\maketitle

\section{\label{intr}Introduction}

Almost two decades have passed from the first evidence for a late-time acceleration of the expansion of the Universe based on type Ia supernovae observations \cite{Perlmutter:1998np,Riess:1998cb}. Since then, more precise cosmological data have confirmed these first results and provided overwhelming evidence for such an acceleration \cite{Suzuki:2011hu,Parkinson:2012vd,Hinshaw:2012aka,Anderson:2012sa,Ade:2015xua}. Most cosmological observations are remarkably consistent with a six parameter spatially flat $\Lambda {\rm CDM}$ model where the late-time dynamics of the Universe is dominated by a cosmological constant $\Lambda$ accounting for nearly $70 \%$ of the total energy density of the Universe.

Despite its successes on the observational side, the $\Lambda {\rm CDM}$ model is faced with yet unsolved fundamental challenges, in particular regarding the small magnitude of $\Lambda$ and the coincidence of the era where it becomes dynamically relevant with the present epoch (see, e. g., \cite{Barreira:2011qi}). Hence, a dynamical scalar field violating the strong energy condition offers a better motivated alternative to the cosmological constant $\Lambda$ in the attempt to explain the current acceleration of the Universe \cite{Copeland:2006wr,Frieman:2008sn,Caldwell:2009ix,Li:2011sd,Bamba:2012cp}. On the other hand, it is reasonable to expect that such field could couple to other fields, possibly leading to measurable variations of nature's fundamental couplings \cite{Carroll:1998zi,Chiba:2001er,Wetterich:2002ic,Nunes:2003ff,Anchordoqui:2003ij,Copeland:2003cv,Parkinson,Doran:2004ek,Marra:2005yt,Avelino:2008dc,Avelino:2009fd,Dent:2008vd,Avelino:2011dh,Avelino:2014xsa}. 

One of such couplings is the fine-structure constant whose dynamics over a wide redshift range is severely constrained using both cosmological, astrophysical and terrestrial data, as well as local laboratory experiments (see \cite{Uzan:2010pm} for a recent review). Despite a few positive claims for a detection of a spacetime variation of the fine-structure constant \cite{Webb:2000mn,Murphy:2003hw,King:2012id}, there is presently no unambiguous  evidence for such variation. Moreover, low redshifts laboratory experiments \cite{Rosenband:2008} (see also \cite{Luo:2011cf} for a recent review of atomic clock constraints on the variation of fundamental couplings) and the Oklo natural nuclear reactor \cite{Gould:2007au,Petrov:2005pu} provide stringent limits on the time variation of $\alpha$.

In \cite{Martins:2015ama,Martins:2015jta,Martins:2016oyv} it has been suggested that upper limits on the time variation of the fine-structure constant tightly constrain the coupling of a dark energy scalar field to the electromagnetic sector, and, indirectly, the violation of the weak equivalence principle. Furthermore, it has also been suggested that such limits, in combination with standard methods, could be used to improve the constraints on the equation of state of dark energy. In this paper we shall demonstrate that such constraints rely on the specification of priors consistent but not favoured by current data. We will  further show that, even in the case of an idealised model where the DE is described by a dynamical scalar field with a constant equation of state parameter $w$ linearly coupled to the electromagnetic sector through a dimensionless coupling $\zeta$, when such priors are relaxed only the combination $|\zeta| {\sqrt {|w+1|}}$ is tightly constrained by current upper limits on the time variation of the fine-structure constant.

Throughout this paper we shall use units with $\hbar=c=8 \pi G=1$ and a metric signature $(+,-,-,-)$.

\section{Dark energy and varying couplings \label{sec2}}

Here we shall assume that the late-time acceleration of the Universe is due to a dynamical dark energy scalar field $\phi$ non-minimally coupled to the electromagnetic field. A particularly interesting class of dark energy models may be defined by the action
\be
S=\int d^4x\,\sqrt{-g}\mathcal\,{\cal L}\,, \label{S1}
\ee
where $g$ is the determinant of the metric tensor, 
\bq
{\cal L} &=& {\cal L}_\phi + {\cal L}_{\phi F} + {\cal L}_{\rm other}\,, \label{S2}\\
{\cal L}_\phi&=&\pm X-V(\phi)\,, \label{S2a}\\
X&=&\pm \frac{1}{2}\partial^\mu \phi \partial_\mu \phi\,, \label{S3}
\eq
$V(\phi)$ is the scalar field potential,
\be 
{\cal L}_{\phi F}= -\frac{1}{4} B_F (\phi)F_{\mu \nu}F^{\mu \nu}\,, \label{S4} 
\ee
$B_F(\phi)$ is a gauge kinetic function, $F_{\mu \nu}$ are the components of 
the electromagnetic field tensor, and ${\cal L}_{\rm other}$ is the Lagrangian density of the other fields. In this class of models the fine-structure constant is given by 
\be 
\alpha(\phi)=\frac{\alpha_0}{B_F(\phi)}\,,
\ee 
where the subscript `$0$' denotes the present time ($B_F(\phi_0)=1$ today). 

\subsection{Time variation of the fine-structure constant}

In the family of models described by Eqs. (\ref{S1}-\ref{S4}) the evolution of $\phi$ induced solely by its coupling to electromagnetically interacting matter is so small, given weak equivalence principle constraints (see  \cite{Uzan:2010pm} and references there in), that the resulting time variation of $\alpha$ can be neglected. Hence, we shall assume that the dynamics of $\phi$ is fully driven by the scalar field potential $V(\phi)$ (and damped by the expansion). On the other hand, since the sound speed of the scalar field $\phi$ equals the speed of light, the spatial variations of the scalar field $\phi$ are small and their contribution to the variation of $\alpha$ may also be neglected in this context.

Consider a flat homogeneous and isotropic Friedmann-Robertson-Walker universe whose dynamics obeys the Friedmann equation given by
\be
H^2 = \frac{\rho}{3}=\frac{\rho_\phi+\rho_{[\phi F+{\rm other}]}}{3}\,, \label{fried} 
\ee
where $\rho_\phi$ is the dark energy density (associated with ${\cal L}_{\phi}$), $\rho_{[\phi F+{\rm other}]}$ is the energy density associated with the remaining lagrangian components (${\cal L}_{\phi F}$ and $ {\cal L}_{\rm other}$), $H={\dot R}/R$ is the Hubble parameter, $R$ is the scale factor, and a dot represents a derivative with respect to the  physical time, $t$. 

Taking into account that the energy density and pressure associated with the scalar field $\phi$ are given respectively by
\be
\rho_\phi=\pm {\dot \phi}^2/2+V(\phi) \,, \quad  p_\phi=\pm {\dot \phi}^2/2-V(\phi)\,, \label{rhop}
\ee
one obtains
\be
w \equiv \frac{p_\phi}{\rho_\phi}= -1 \pm  \frac{{\phi'}^2H^2}{\rho_\phi}=-1 \pm \frac{{\phi'}^2}{3\Omega_\phi}\,, \label{wphi}
\ee
where $\Omega_\phi=\rho_\phi/\rho$ and a prime represents a derivative with respect to $\ln R$ (${\dot \phi}=\phi' H$).

If the gauge kinetic function $B_F$ is a linear function of $\phi$ with $|B_F(\phi)-1| \ll 1$ then
\be 
\frac{\dot \alpha}{\alpha}=\zeta {\dot \phi}\,, \label{linear}
\ee 
where $\zeta$ is a constant. In the following we shall only consider the solution with $\dot \phi > 0$, so that the sign of $\dot \alpha$ is the same as that of $\zeta$.  Note, however, that this assumption may be relaxed since both  $ {\cal L}_{\phi}$ and ${\cal L}_{\phi F}$ are invariant under the transformation $\phi \to -\phi$, $V(\phi) \to V(-\phi)$, $\zeta \to -\zeta$.

Eqs. (\ref{wphi}) and (\ref{linear}) imply that
\be
\frac{1}{H}\frac{\dot \alpha}{\alpha} =\zeta \sqrt{3\Omega_\phi(z)|1+w|}\,. \label{alphadot}
\ee

\subsubsection{Constant $w$ models}

The time evolution of the energy density associated with the dark energy scalar field $\phi$ obeys the equation
\be
{\dot \rho_\phi}+3H(1+w)=0\,. \label{adiabatic}
\ee
In this paper, for simplicity, we shall consider a constant equation of state parameter $w$ smaller than unity (see \cite{Avelino:2009ze,Avelino:2011ey} for a discussion of constant $w$ models) --- relaxing this assumption would only strengthen our conclusions. Then, from Eqs. (\ref{rhop}), (\ref{wphi}) and (\ref{adiabatic}), it is possible to show that
\be
{\dot \phi}=\left(2V\frac{|1+w|}{1-w}\right)^{1/2}\,.
\ee
with
\be
V=V_0 R^{-3(1+w)}\,,
\ee
where the scale factor at the present time is normalised to unity ($R_0=1$).
Together with Eq. (\ref{linear}) this implies that
\be 
\frac{\dot \alpha}{\alpha}=\zeta {\dot \phi} \propto (1+z)^{3(1+w)/2}\,, \label{alphazev}
\ee 
has a very slow evolution evolution with the redshift $z$ ($1+z=1/R$) for values of $w$ close to $-1$ consistent with the current observations. 

\subsection{Constraints on the time variation of the fine-structure constant}

In this paper, for simplicity, we shall focus on the atomic clock constraint on the present variation of $\alpha$ with time of Rosenband {\it et al.}  \cite{Rosenband:2008}
\be
\left. \frac{\dot\alpha}{\alpha} \right |_{z=0} =(-1.6\pm2.3)\times10^{-17}\,{\rm yr}^{-1}\,. \label{Rosen}
\ee
which is currently the strongest laboratory constraint on $\alpha$ alone. Eq. (\ref{alphazev}) implies that, for a constant $w$, this local laboratory constraint on the value of $\alpha$ is significantly more constraining than most other astrophysical and cosmological constraints (in particular than the astrophysical constraints on the variation of $\alpha$ considered in \cite{Martins:2015ama,Martins:2015jta,Martins:2016oyv}). 

The constraint given in Eq. (\ref{Rosen}) may be rewritten in a dimensionless form as
\begin{equation} 
\left. \frac{1}{H_0}\frac{\dot\alpha}{\alpha} \right |_{z=0} =(-2.3\pm3.3)\times10^{-7}\,,
\end{equation}
taking into account that $H_0=100 \, h \, {\rm km \, s^{-1} \, Mpc^{-1}}$, with $h=0.678$ \cite{Ade:2015xua}, and neglecting the small uncertainty on the current value of the Hubble parameter $H_0$. Assuming a value of $\Omega_{\phi 0} = 0.692$ consistent with the Planck 2015 results \cite{Ade:2015xua} and neglecting the relatively small error bar associated with $\Omega_{\phi 0}$, one finally finds (using Eq. (\ref{alphadot}) evaluated at the present time) that
\begin{equation} 
x = \zeta {\sqrt {|1+w|}}=(-1.6\pm 2.3)\times10^{-7}\,. \label{limitx}
\end{equation}

Eq. (\ref{limitx}) implies that a nontrivial lower limit on the value of $|1+w|$ ($|1+w| \ge |1+w|_{\rm min}>0$), if it existed, could be used to obtain an upper bound on the value of $|\zeta|$,
\begin{equation} 
|\zeta| \le \frac{3.9}{\sqrt {|1+w|_{\rm min}}} \times10^{-7}\,.
\end{equation}
Analogously, Eq. (\ref{limitx}) combined with a nontrivial lower limit on $|\zeta|$ would imply an upper bound on the value of $|1+w|$ associated to constraints on the time evolution of $\alpha$. Nevertheless, there is currently no unambiguous observational evidence favouring a nontrivial lower limit on either $|w+1|$ or $|\zeta|$. As we shall demonstrate in the following section this precludes the use of constraints on the time variation of $\alpha$ to set realistic upper bounds on $|\zeta|$ or $|w+1|$. 

Alternatively, as recognised in \cite{Dvali:2001dd,Chiba:2001er}, a nontrivial lower bound on the value of $|x|$ together with an upper bound on the value of $|w+1|$ could be used to obtain a nontrivial lower bound on $|\zeta|$. However, presently there is also no unambiguous nontrivial lower bound on the value of $|x|$ and, consequently, no non-trivial lower bound on $|\zeta|$ from varying $\alpha$ constraints.

Although in \cite{Martins:2015ama,Martins:2015jta,Martins:2016oyv} cosmological constraints on $|w+1|$ from Type Ia supernova data were considered, here we shall simplify the analysis and account for standard constraints on the value of $|w+1|$ by incorporating them in the $|w+1|$ prior (except if stated otherwise, we shall conservatively assume that $|w+1| \le 0.1$ \cite{Ade:2015xua}).

\section{Role of priors \label{sec3}}

We label the random variables associated with the parameters $\theta={\sqrt {|1+w|}} \ge 0$,  $\zeta$, and $x=\theta \zeta$ by $\Theta$, $Z$, and $X=\Theta Z$, respectively. We shall investigate the impact of the prior on the random variable $\Theta$ on the estimation of $|Z|$, assuming, for simplicity, that $\Theta$ and $X$ are independent random variables. This means that the prior on $\Theta$ is the same as the posterior, since, in this case, the probability density function for the variable $\Theta$ is not altered by a measurement of $X$. This assumption allows us to derive analytical expressions for the probability density function of $|Z|$ for various priors of $\Theta$, but does not otherwise affect our main results.

Given that $\Theta$ and $X$ are assumed to be independent, the cumulative distribution function of the random variable $|Z|=|X|/\Theta$ is given by
\bq
F_{|Z|}(|\zeta|) &\equiv& P(|Z| \le |\zeta|)= \nonumber \\ 
& &\int_0^\infty \left( \int_0^{\theta |\zeta|}  f_{|X|} (|x|)  f_{\Theta} (\theta)  d |x| \right) d \theta\,, \label{cdis}
\eq
so that the corresponding probability density function is
\be
f_{|Z|}(|\zeta|)=\frac{F_{|Z|}(|\zeta|)}{d|\zeta|} = \int_0^\infty \theta f_{|X|} (\theta|\zeta|)  f_{\Theta} (\theta) d \theta\,. \label{fZ}
\ee

We shall consider a probability density function for the random variable $|X|$,
\bq
f_{|X|}(|x|)&=&\frac{1}{\sqrt {2 \pi} \sigma} \left(\exp \left[-\frac{(|x|-\mu)^2}{2\sigma^2}\right]+\right. \nonumber \\
& &\left. \exp \left[-\frac{(|x|+\mu)^2}{2\sigma^2}\right]\right)\,, \label{fX}
\eq
consistent with the atomic clock constraint on the present variation of $\alpha$ with time of Rosenband {\it et al.}  \cite{Rosenband:2008} discussed in the previous section ($\mu=-1.6 \times 10^{-7}$ and $\sigma=2.3 \times 10^{-7}$).

\begin{figure}
\includegraphics[width=3.4in]{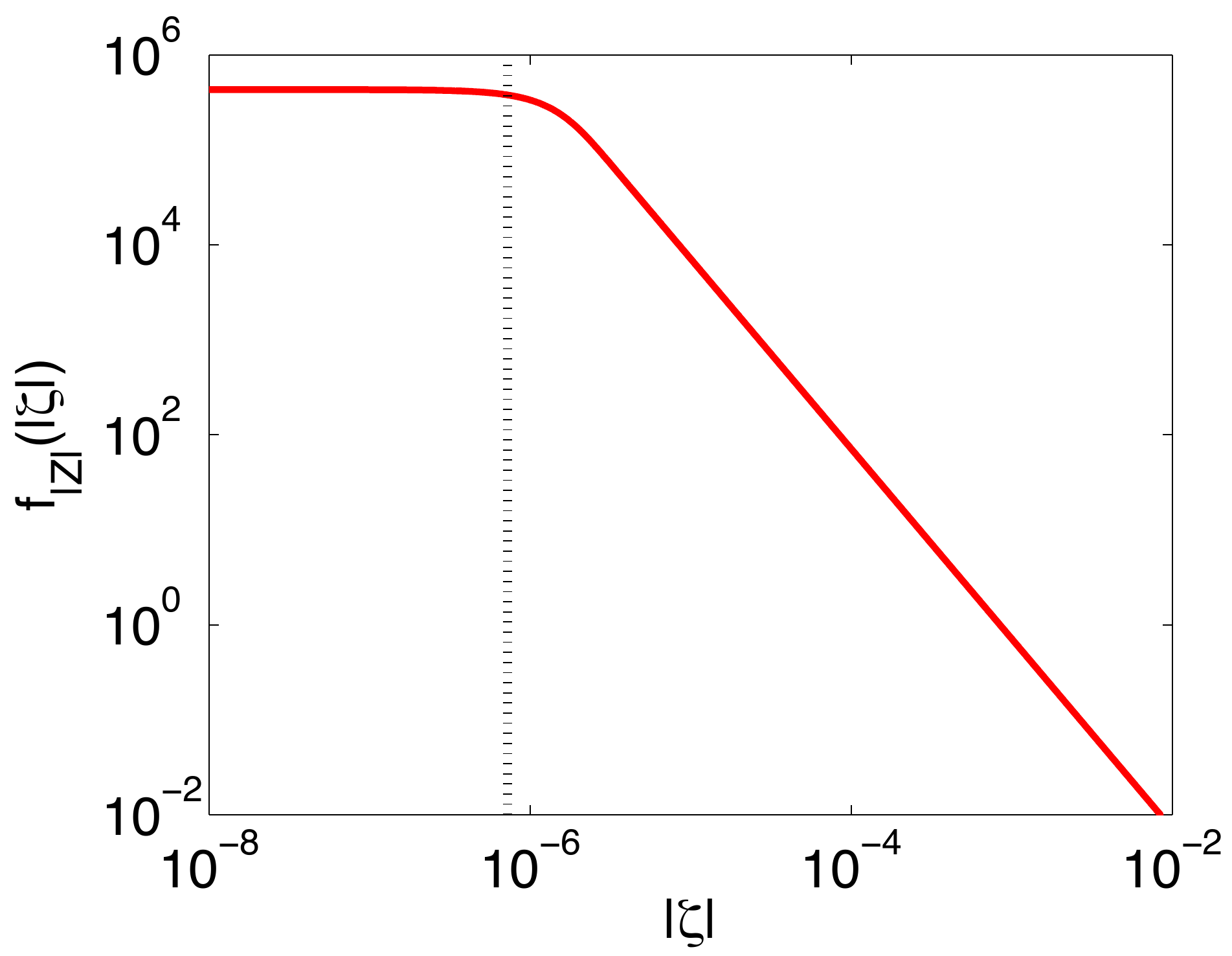}
\caption{The red solid line represents the probability density function $f_{|Z|}(|\zeta|)$ given by Eqs. (\ref{fu1}) and  (\ref{fu2}) with $\mu=-1.6 \times 10^{-7}$, $\sigma=2.3 \times 10^{-7}$ and $b \equiv {\sqrt {|w+1|}_{\rm max}}=10^{-1/2}$. The vertical dotted line is defined by $\zeta_b \equiv \sigma/b$. $f_{|Z|}$ is nearly constant for $|\zeta| < \zeta_b$ and decays roughly proportionally to $\zeta^{-2}$ for $|\zeta| > \zeta_b$.}
\label{Fig1}
\end{figure}

\subsection{Uniform prior}

Let us start by assuming that the probability density function of the variable $\Theta$ is uniform in the interval $[0,b]$ and vanishes outside it, so that
\be
f_{\Theta}(\theta)=\frac{1}{b}\,, \qquad 0 \le \theta \le b\,,
\ee
where $b \equiv {\sqrt {|w+1|}_{\rm max}}$.

In this case, the probability density function of the variable $|Z|$ may be computed analytically, using Eqs. (\ref{cdis}) and (\ref{fZ}), and it is given by
\be
f_{|Z|}(|\zeta|)=f^+_{|Z|}(|\zeta|)+f^-_{|Z|}(|\zeta|)\,, \label{fu1}
\ee
where
\bq
f^\pm_{|Z|}(|\zeta|)&=&\frac{1}{{\sqrt {8 \pi}} b |\zeta|^2} \left(2\sigma\left[\exp\left(-\frac{\mu^2}{2\sigma^2}\right)-\right. \right. \nonumber \\
& &\left. \exp\left(-\frac{(\mu \pm b|\zeta|)^2}{2\sigma^2}\right)\right]+  \nonumber \\
& &\left. {\sqrt {2 \pi}} \mu \left( {\rm erf}\left[\frac{\mu}{{\sqrt 2}\sigma}\right]-{\rm erf}\left[\frac{\mu \pm b|\zeta|}{{\sqrt 2}\sigma}\right] \right)\right)\,, \label{fu2}
\eq
and the error function is defined by
\be
{\rm erf}(y) \equiv \frac{2}{\sqrt \pi} \int_0^y e^{-u^2} du\,.
\ee

\begin{figure}
\includegraphics[width=3.4in]{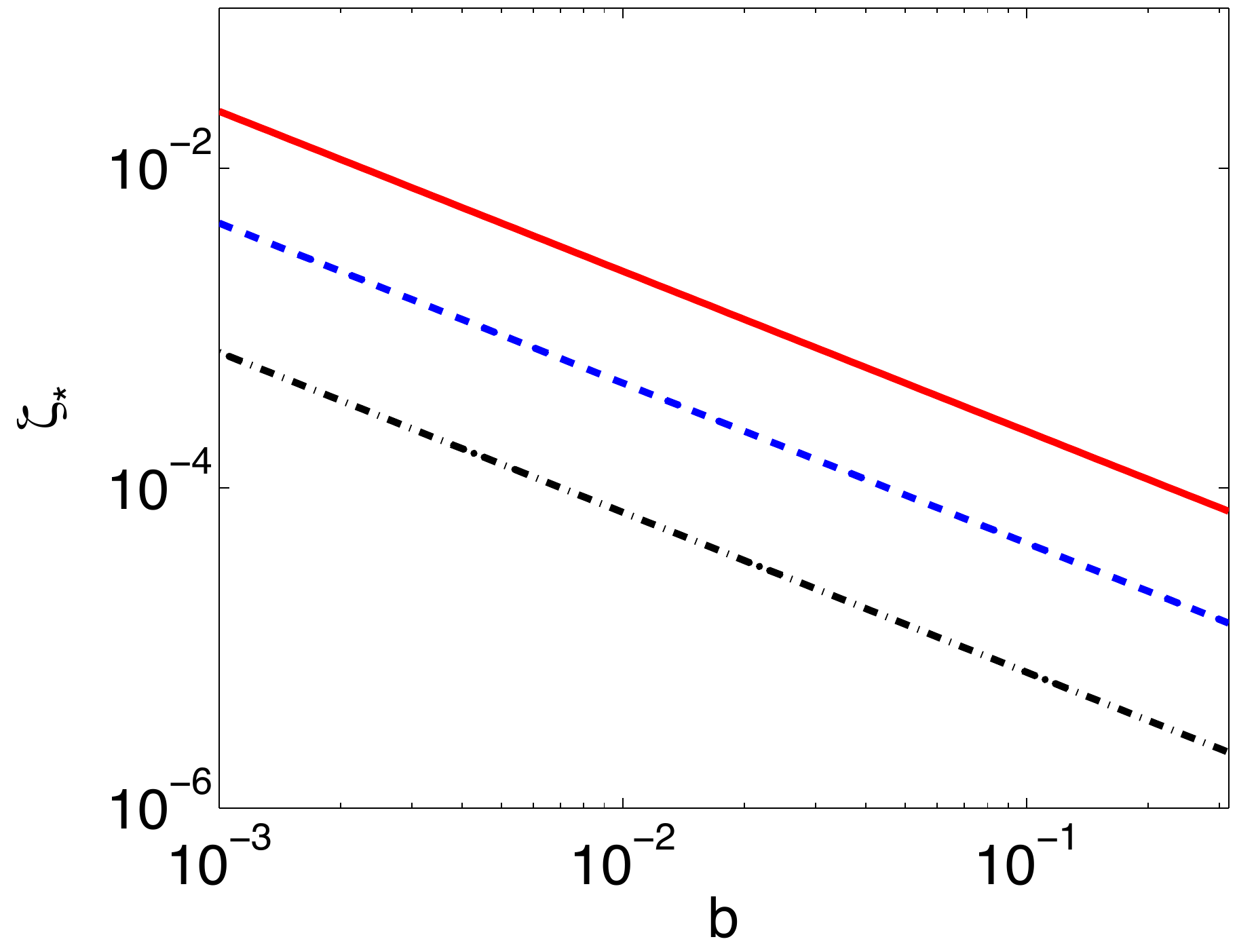}
\caption{The value of $\zeta_*$ such that $F_{|Z|}(\zeta_*) \equiv P(|Z| \le \zeta_*) =\chi$ (calculated using Eqs. (\ref{fu1}) and  (\ref{fu2}) with $\mu=-1.6 \times 10^{-7}$, $\sigma=2.3 \times 10^{-7}$) as a function of $b \equiv {\sqrt {|w+1|}_{\rm max}}$, for $\chi=0.99$ (red solid line), $\chi=0.95$ (blue dashed line) and $\chi=0.68$ (black dot-dashed line).}
\label{Fig2}
\end{figure}

Fig. 1 shows $f_{|Z|}$, given by Eqs. (\ref{fu1}) and  (\ref{fu2}) with $\mu=-1.6 \times 10^{-7}$, $\sigma=2.3 \times 10^{-7}$ and $b \equiv {\sqrt {|w+1|}_{\rm max}}=10^{-1/2}$, as a function of $|\zeta|$ (red solid line). The vertical dotted line is defined by $\zeta_b \equiv \sigma/b$. The probability density function $f_{|Z|}$ is nearly constant for $|\zeta| < \zeta_b$ and decays roughly  proportionally to $\zeta^{-2}$ for $|\zeta| > \zeta_b$. Fig. 2 shows the value of $\zeta_*$ such that $F_{|Z|}(\zeta_*) \equiv P(|Z|  \le  \zeta_*)=\chi$ (calculated using Eqs. (\ref{fu1}) and  (\ref{fu2}) with $\mu=-1.6 \times 10^{-7}$, $\sigma=2.3 \times 10^{-7}$) as a function of $b \equiv {\sqrt {|w+1|}_{\rm max}}$, for $\chi=0.99$ (red solid line), $\chi=0.95$ (blue dashed line) and $\chi=0.68$ (black dot-dashed line).  Fig. 2 shows that a weaker prior on $|\Theta|$ leads to stronger constraints on $|Z|$. This is directly associated with the choice of a uniform prior in the interval $[0,b]$ which disfavours very small values of $|\zeta|$, specially if $b$ is large. Also note that, due to the heavy tail of $f_{|Z|}$, the constraints on $|Z|$ degrade very rapidly as one increases the confidence level $\chi$. For $b \equiv {\sqrt {|w+1|}_{\rm max}}=10^{-1/2}$ and $\chi=0.68$ one obtains $\zeta_*=2.3 \times 10^{-6}$, which is in reasonable agreement with the results obtained in \cite{Martins:2015ama,Martins:2015jta,Martins:2016oyv} considering a flat prior for the equation of state parameter of the dark energy.

\subsection{Logarithmic prior}

Let us now consider the case of a uniform probability density function of the variable $\ln \Theta$ for $\theta$ in the interval $[a,b]$, with $0 < a < b$. The corresponding probability density function of the variable $\Theta$ is 
\be
f_{\Theta}(\theta)=\left[\ln \frac{b}{a}\right]^{-1} \frac{1}{\theta}\,, \qquad a \le \theta \le b\,,
\ee
and it is equal to zero outside the interval $[a,b]$.
In this case, the probability density function of the variable $|Z|$ may also be computed analytically, using Eqs. (\ref{cdis}) and (\ref{fZ}), and it is given by
\bq
& &f_{|Z|}(|\zeta|)= \left[\ln \frac{b}{a}\right]^{-1} \frac{1}{2|\zeta|} \left({\rm erf}\left[\frac{\mu+b|\zeta|}{{\sqrt 2}\sigma}\right]-\right. \nonumber \\
& &\left. {\rm erf}\left[\frac{\mu-b|\zeta|}{{\sqrt 2}\sigma}\right]-{\rm erf}\left[\frac{\mu+a|\zeta|}{{\sqrt 2}\sigma}\right]+{\rm erf}\left[\frac{\mu-a|\zeta|}{{\sqrt 2}\sigma}\right]\right)\,. \label{fl}
\eq

\begin{figure}
\includegraphics[width=3.4in]{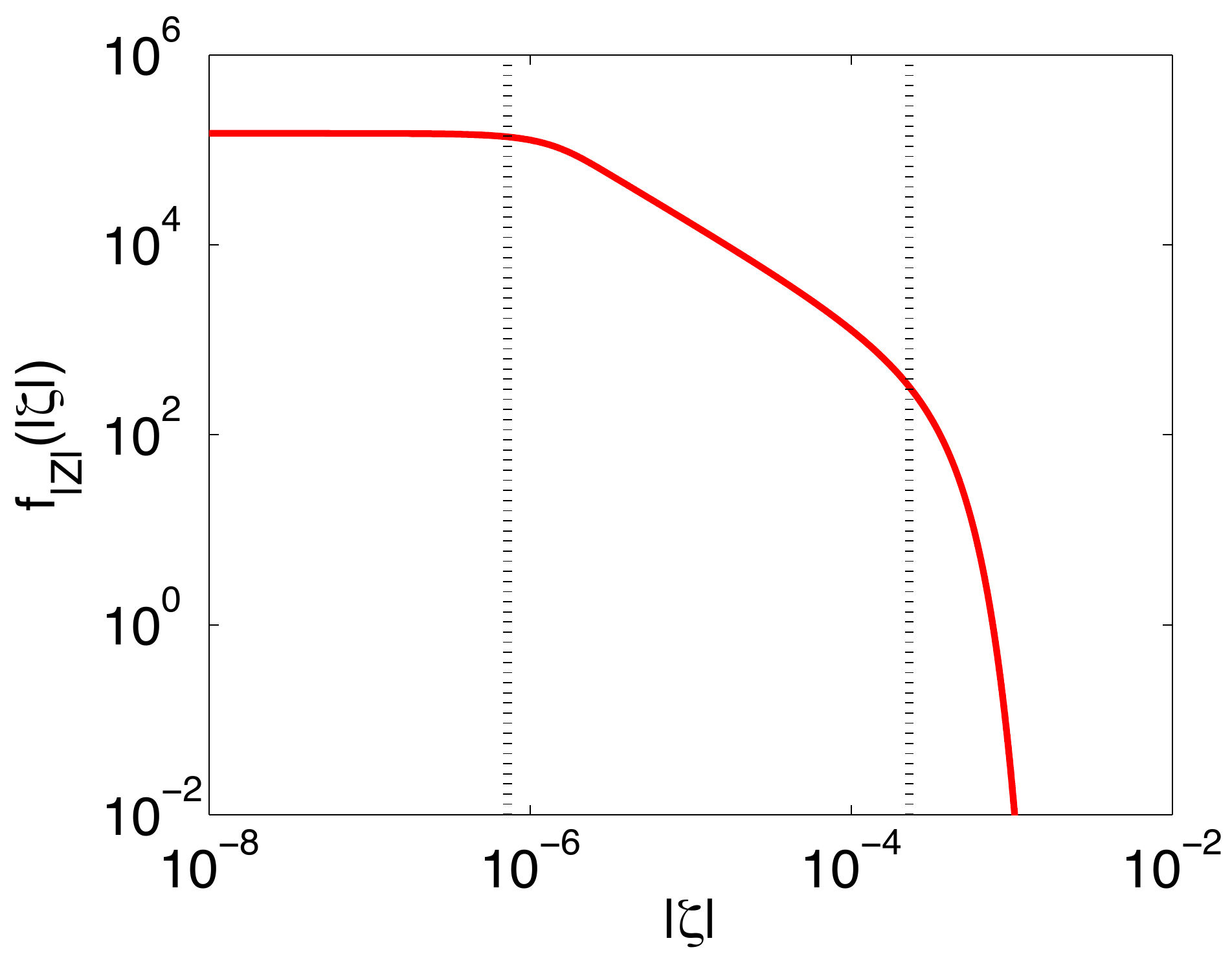}
\caption{The red solid line represents the probability density function $f_{|Z|}(|\zeta|)$ given by Eq. (\ref{fl}) with $\mu=-1.6 \times 10^{-7}$, $\sigma=2.3 \times 10^{-7}$, $a \equiv {\sqrt {|w+1|}_{\rm min}}=10^{-3}$ and $b \equiv {\sqrt {|w+1|}_{\rm max}}=10^{-1/2}$. The vertical dotted lines are defined by $\zeta_b \equiv \sigma/b$ and $\zeta_a \equiv \sigma/a$ (left and right dotted lines, respectively). $f_{|Z|}(|\zeta|)$  is nearly constant for $|\zeta| < \zeta_b$ and decays roughly proportionally to $\zeta^{-1}$ for $\zeta_b < |\zeta| < \zeta_a$ (and much faster than that for $|\zeta| > \zeta_a$).}
\label{Fig3}
\end{figure}

Fig. 3 shows $f_{|Z|}$, given by Eq. (\ref{fl}) with $\mu=-1.6 \times 10^{-7}$, $\sigma=2.3 \times 10^{-7}$, $a \equiv {\sqrt {|w+1|}_{\rm min}}=10^{-3}$ and $b \equiv{\sqrt {|w+1|}_{\rm max}}=10^{-1/2}$, as a function of $|\zeta|$ (red solid line). The vertical dotted lines are defined by $\zeta_b \equiv \sigma/b$ and $\zeta_a \equiv \sigma/a$ (left and right dotted lines, respectively). The probability density function is nearly constant for $|\zeta| < \zeta_b$ and decays roughly proportionally to $\zeta^{-1}$ for $\zeta_ b < |\zeta| <  \zeta_a$ (and much faster than that for $|\zeta| > \zeta_a$). Fig. 4 shows the value of $\zeta_*$ such that $F_{|Z|}(\zeta_*) \equiv P(|Z| \le |\zeta_*|) =\chi$ (calculated using Eq. (\ref{fl}) with $\mu=-1.6 \times 10^{-7}$, $\sigma=2.3 \times 10^{-7}$ and $b \equiv {\sqrt {|w+1|}_{\rm max}}=10^{-1/2}$), as a function of $a$, for $\chi=0.99$ (red solid line), $\chi=0.95$ (blue dashed line) and $\chi=0.68$ (black dot-dashed line).  Fig. 2 shows that the lower the value of $a$, the weaker the constraints on $|Z|$ become. For $a \equiv {\sqrt {|w+1|}_{\rm min}} = 10^{-2}$ and $\chi=0.68$ one obtains $\zeta_*=6.0 \times 10^{-6}$, in reasonable agreement with the results obtained in \cite{Martins:2016oyv} using a logarithmic prior for the equation of state parameter of the dark energy. However, in the $a \equiv {\sqrt {|w+1|}_{\rm min}} \to 0$ limit the logarithmic prior favours values of $|w+1|$ extremely close to zero, which is the reason why, in this limit, the constraints on $|Z|$ become extremely weak. Again, note the rapid degradation of the constraints on $|Z|$ with the increase of the confidence level $\chi$.

\begin{figure}
\includegraphics[width=3.4in]{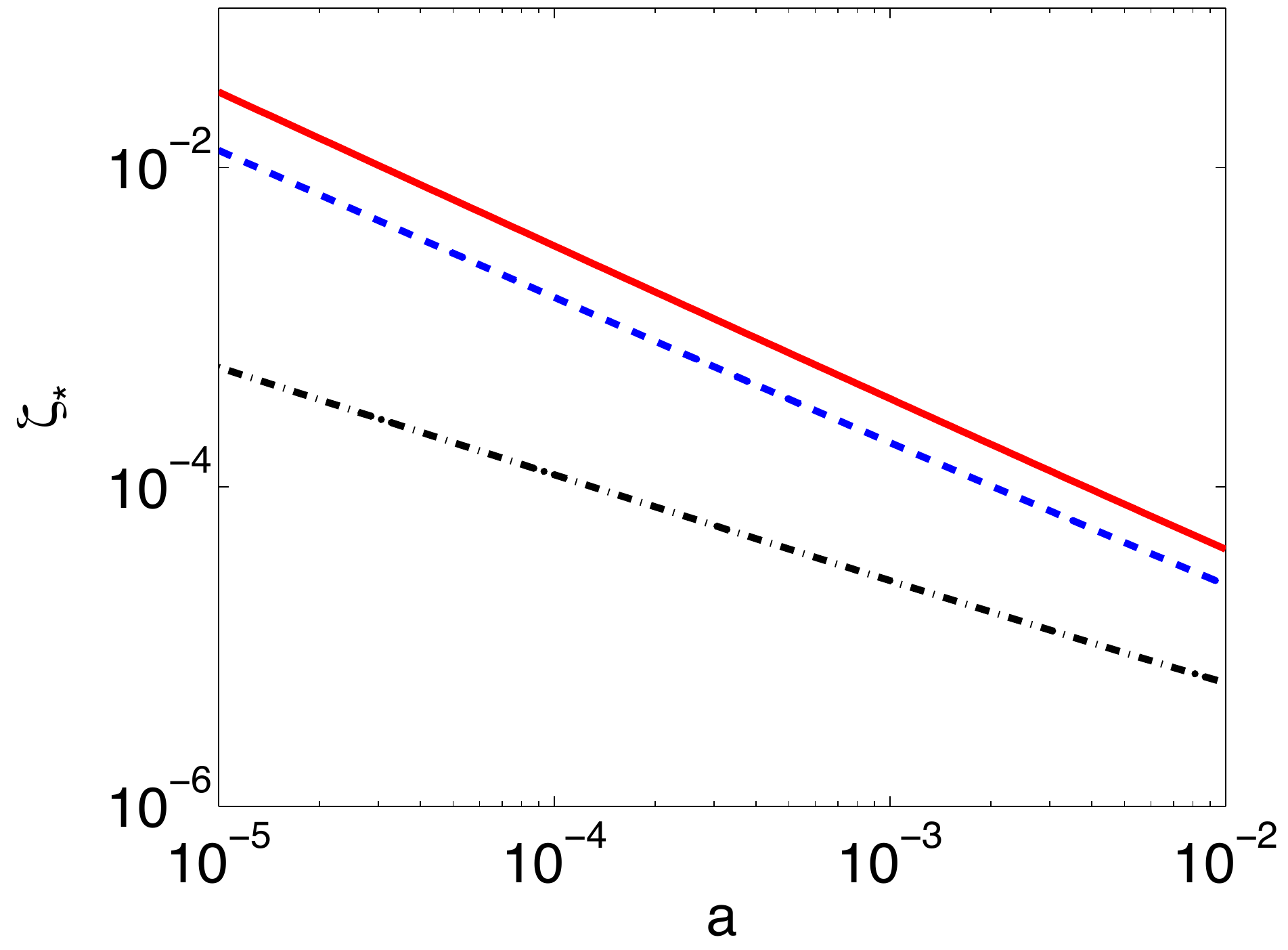}
\caption{The value of $\zeta_*$ such that $F_{|Z|}(\zeta_*) \equiv P(|Z| \le \zeta_*) =\chi$ (calculated using Eq. (\ref{fl}) with $\mu=-1.6 \times 10^{-7}$, $\sigma=2.3 \times 10^{-7}$ and $b \equiv {\sqrt {|w+1|}_{\rm max}}=10^{-1/2}$), as a function of $a \equiv {\sqrt {|w+1|}_{\rm min}}$, for $\chi=0.99$ (red solid line), $\chi=0.95$ (blue dashed line) and $\chi=0.68$ (black dot-dashed line).}
\label{Fig4}
\end{figure}

\begin{figure}
\includegraphics[width=3.4in]{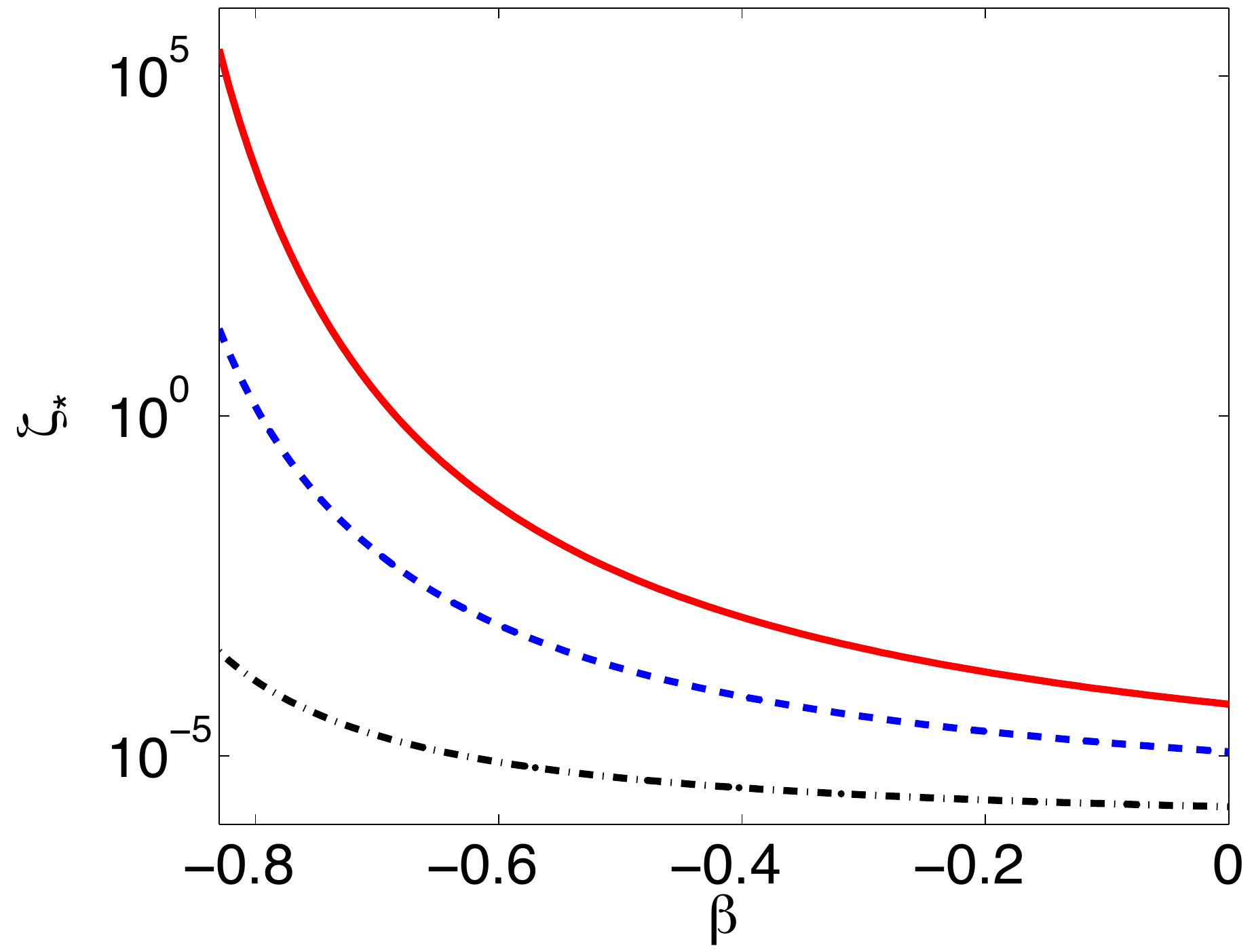}
\caption{The value of $\zeta_*$ such that $F_{|Z|}(\zeta_*) \equiv P(|Z| \le \zeta_*) =\chi$ (calculated using Eq. (\ref{fZpower}) with $\sigma=2.3 \times 10^{-7}$ and $b \equiv {\sqrt {|w+1|}_{\rm max}}=10^{-1/2}$), as a function of $\beta$, for $\chi=0.99$ (red solid line), $\chi=0.95$ (blue dashed line) and $\chi=0.68$ (black dot-dashed line).}
\label{Fig5}
\end{figure}

\subsection{Power law prior}

A more general class of probability density functions, which includes uniform probability density functions for the variable $\Theta$ as a particular sub-class and uniform probability density functions for the variable $\ln \Theta$ as a special limit for $\beta \to -1$, is given by
\be
f_{\Theta}(\theta)=\frac{\beta+1}{b} \left(\frac{\theta}{b}\right)^\beta \,, \qquad \theta \le b\,,
\ee
with $f_{\Theta}(\theta)=0$ for $\theta > b$ (here, $\beta > -1$). Taking the probability density function for the variable $X$ given in Eq. (\ref{fX}), but now assuming $\mu=0$, $f_{|Z|}(|\zeta|)$ may be computed analytically using Eqs. (\ref{cdis}) and (\ref{fZ}). The result is 
\bq
f_{|Z|}(|\zeta|)&=& \frac{b(\beta+1)}{\sqrt {2 \pi} \sigma} \left(\frac{b^2 \zeta^2}{2 \sigma^2}\right)^{-\beta/2-1} \times \nonumber \\
& &\left( \Gamma\left[\frac{\beta}{2}+1\right]-\Gamma\left[\frac{\beta}{2}+1,\frac{b^2 \zeta^2}{2 \sigma^2}\right]\right)\,, \label{fZpower}
\eq
where
\be
\Gamma(y) \equiv \int_0^\infty u^{y-1} e^{-u} du
\ee
is the Gamma function and $\Gamma(y,z)$ is the upper incomplete Gamma function defined by
\be
\Gamma(y,z) \equiv \int_z^\infty u^{y-1} e^{-u} du\,.
\ee
The value of $\mu$ implied by Eq. (\ref{limitx}) is less than one sigma away from zero. Consequently, the error committed in assuming that $\mu=0$ is relatively small, thus justifying the use of this approximation in order to obtain the analytical result for $f_{|Z|}(|\zeta|)$ given by Eq. (\ref{fZpower}).

Fig. 5 shows the value of $\zeta_*$ such that $F_{|Z|}(\zeta_*) \equiv P(|Z| \le \zeta_*) =\chi$ (calculated using Eq. (\ref{fZpower}) with $\sigma=2.3 \times 10^{-7}$ and $b \equiv {\sqrt {|w+1|}_{\rm max}}=10^{-1/2}$), as a function of $\beta$, for $\chi=0.99$ (red solid line), $\chi=0.95$ (blue dashed line) and $\chi=0.68$ (black dot-dashed line). It shows that for values of $\beta$ sufficiently close to $-1$ the constraints on $|Z|$ become extremely weak. Notice that the the $68\%$, $95\%$ and $99\%$ constraints on $|Z|$, may span several orders of magnitude, in particular for $\beta$ close to $-1$.

\section{\label{conc} Conclusions}

In this paper we critically assessed recent claims suggesting that upper limits on the time variation of $\alpha$ could be used to tightly constrain the dynamics a dark energy scalar field (in particular, its coupling to the electromagnetic sector). We have shown that such constraints rely on assumptions which are consistent but not favoured by current data. This situation could be improved if there was i) a nontrivial lower bound on the value of $|x|$ or ii) a nontrivial lower bound on the value of $|w+1|$ or $|\zeta|$. Although i) may in principle be accomplished with a new generation of high-resolution ultra-stable spectrographs, such as ESPRESSO and ELT-HIRES, and ii) may, in principle, be achieved respectively by forthcoming missions to map the geometry of the Universe, such as Euclid, or to test the equivalence principle, such as MICROSCOPE or STEP, there is a priori no guarantee that these missions will make a detection rather than significantly improving current bounds. In the later case the analysis reported in the present paper will remain pertinent, despite the improved constraints. As demonstrated in \cite{Avelino:2008dc,Avelino:2011dh,Avelino:2014xsa}, even in the more optimistic case of a significant detection, the dependence of dark energy constraints from the time variation of $\alpha$ on crucial assumptions, including i) that general relativity provides an accurate description of gravity on cosmological scales ii) that dark energy may be described by a dynamical scalar field obeying Eqs. (\ref{S1}-\ref{S4})  iii) that the coupling between the dark energy scalar field and the electromagnetic field is linear, should not neglected.

\begin{acknowledgments}

The author thanks Margarida Cunha for enlightening discussions. This work was supported by Funda{\c c}\~ao para a Ci\^encia e a Tecnologia (FCT) through the Investigador FCT contract of reference IF/00863/2012 and POPH/FSE (EC) by FEDER funding through the program Programa Operacional de Factores de Competitividade - COMPETE. Funding of this work was also provided by the FCT Grant No. UID/FIS/04434/2013.
\end{acknowledgments}


\bibliography{alpha}

\end{document}